# Modeling diffusion in ionic, crystalline solids with internal stress gradients

Benjamin L. Hess[a*] and Jay. J. Ague[a,b]

[a] Department of Earth and Planetary Sciences, Yale University, PO Box 208109, New Haven, CT 06520-8109, USA.

[b] Yale Peabody Museum of Natural History, Yale University, New Haven, CT 06511, USA

[*] Corresponding author: Benjamin.hess@yale.edu

**Abstract:**

Intracrystalline diffusion is an invaluable tool for estimating timescales of geological events. Diffusion is typically modeled using gradients in chemical potential caused by variations in composition. However, chemical potential is derived for uniform pressure and temperature conditions and therefore cannot be used to model diffusion when there are gradients in stress. Internal stress variations in minerals will create gradients in strain energy which, in addition to gradients in composition, will drive diffusion. Consequently, it is necessary to have a method that incorporates stress variations into diffusion models.

To address this issue, we have derived a flux expression that allows diffusion to be modeled in ionic, crystalline solids under arbitrary stress states. Our approach is consistent with standard petrological methods but instead utilizes gradients in a thermodynamic potential called "relative chemical potential." Relative chemical potential accounts for the lattice constraint in crystalline solids by quantifying changes in free energy due to the exchanges of constituents on lattice sites under arbitrary stress conditions. Consequently, gradients in relative chemical potential can be used to model diffusion when pressure is not uniform (i.e., under conditions of non-hydrostatic stress).

We apply our derivation to the common quaternary garnet solid solution almandine–pyrope–grossular–spessartine. The rates and directions of divalent cation diffusion in response to stress are determined by endmember molar volume or lattice parameters, elastic moduli, and non-ideal activity interaction parameters. Our results predict that internal stress variations of one hundred MPa or more are required to shift garnet compositions by at least a few hundredths of a mole fraction. Mineral inclusions in garnet present a potential environment to test and apply our stress-driven diffusion approach, as stress variations ranging from hundreds of MPa to GPa-level are observed or predicted around such inclusions. The ability to model stress-induced diffusion may provide new information about the magnitudes of both intracrystalline stresses and the timescales during which they occurred, imparting a better understanding of large-scale tectono-metamorphic processes.

**1. Introduction**

Intracrystalline diffusion is invaluable for inferring temperature–time histories of lithospheric processes. For example, diffusion profiles between compositional zones within garnet or at garnet rims in contact with matrix phases have been used to determine the timescales of metamorphic events including rates of prograde heating, fault movement, fluid infiltration, retrograde cooling, and exhumation (e.g., Erambert and Austrheim, 1993; Florence and Spear, 1995; Van Orman et al., 2001; Perchuk, 2002; Cherniak and Watson, 2003; Faryad and Chakraborty, 2005; Carlson, 2006; Ague and Baxter, 2007; Raimbourg et al., 2007; Gaidies et al., 2008; Caddick et al., 2010; Ganguly, 2010; Vorhies and Ague, 2011; Viete et al., 2011, 2018; Chu et al., 2017, 2018; Tan et al., 2020; Zou et al., 2021).

However, in many of these tectonic processes, significant variations in stress (i.e., non-hydrostatic stress) can exist between and within mineral grains (e.g., Wheeler, 1987, 2014, 2020; Moulas et al., 2013; Tajčmanová et al., 2014, 2015; Zhong et al., 2017, 2019; Jamtveit et al., 2019; Moore et al., 2019; Wallis et al., 2019, 2022), which presents a potential problem for modeling diffusion. Standard diffusion models are based on gradients in the Gibbsian chemical potential of each endmember (e.g., Loomis, 1978; Zhang, 2010; Borinski et al., 2012; Yoo,

2020). However, chemical potential is derived for conditions of uniform pressure (i.e., hydrostatic stress; Gibbs, 1878). As such, chemical potential gradients (and corresponding mole fraction gradients) alone cannot be used to model diffusion in the presence of stress gradients (Larché and Cahn, 1982; Cahn and Larché, 1983).

To overcome this challenge, materials scientists F. C. Larché and J. W. Cahn (1973, 1985) used the calculus of variations approach following Gibbs (1878) to derive a new thermodynamic potential that can be used to both model diffusion and determine equilibrium compositions in multi-component solids under arbitrary stress conditions. We refer to this potential as relative chemical potential after Gurtin et al. (2010, p. 399). Relative chemical potential applies to any solid that can be described by a defined network that exists independently of the constituents which fill the network, as is the case for crystalline solids or glasses. Example constituents within networks include metal atoms in alloys and ions in silicates (Larché and Cahn, 1973, 1985). Relative chemical potential defines the energy difference due to the *exchange* of two constituents (e.g., atoms or ions) on a network site (e.g., a crystalline lattice site) while holding all other constituents constant (Larché and Cahn, 1973, 1985).

Although the chemical potential cannot be defined within the interior of a solid for unequal (non-hydrostatic) stress conditions (Gibbs, 1878; Kamb, 1961; Larché and Cahn, 1985; Wheeler, 2018; Hess et al., 2022), the change in energy due to the exchange of constituents within a solid, as governed by the relative chemical potential, can be (Larché and Cahn, 1973, 1985). Diffusion will occur until the overall energy cannot be lowered in the solid by further exchanges of constituents on the sites (Larché and Cahn, 1973, 1982, 1985; Voorhees and Johnson, 2004; Powell et al., 2018, 2019; Wheeler, 2018; Hess et al., 2022). Diffusion following the relative chemical potential gradient will lead to positive entropy production as required for any real process (Hess et al., 2022). Thus, the relative chemical potential can be used to treat diffusion in solids with internal stress gradients. Larché-Cahn theory has been experimentally verified in alloys near room temperature and pressure (Shi et al., 2018). While it remains untested in crystals at high temperatures and pressures, we posit that the strong silicate structures present in many minerals are consistent with the network model, and thus the relationships between stress and solid chemistry derived by Larché and Cahn (1973, 1985) are applicable.

It is highly desirable to quantitatively evaluate diffusion under stress gradients given the increasing awareness of appreciable grain-scale stress variation (e.g., Neusser et al., 2012; Tajčmanová et al., 2014, 2015; Wheeler, 2014, 2018, 2020; Zhong et al., 2017, 2019; Thomas and Spear, 2018; Bonazzi et al., 2019; Jamtveit et al., 2019; Moore et al., 2019; Wallis et al., 2019, 2022; Campomenosi et al., 2020; Hess and Ague, 2021; Hess et al., 2022). Important progress has been made in this regard by modeling coupled stress, deformation, and diffusion for binary systems without the lattice constraint (e.g., Zhong et al., 2017). However, a general treatment for diffusion in ionic, crystalline solids with any number of endmembers that is consistent with the lattice constraint and Larché-Cahn theory remains to be formulated.

In this work we provide a method for modeling diffusion that 1) can incorporate gradients in stress, 2) results in the equilibrium conditions predicted by Larché-Cahn theory (Larché and Cahn, 1985), and 3) is based on self-diffusion coefficients widely used in petrology for multi-component ionic solid solutions with any number of endmembers. We accomplish this using an approach which begins with the same assumptions as the standard methods used for modeling diffusion in ionic systems (Wendt 1965; Lasaga, 1979) and incorporating the definition of relative chemical potential (Larché and Cahn, 1973). The resulting flux expression is consequently valid for modeling diffusion in ionic, multi-component minerals under arbitrary stress conditions. Diffusion in almandine–pyrope–grossular–spessartine (Alm–Py–Grs–Sps) garnet is then explored as an example.

## 2. Methods

### 2.1. Review of standard uniform pressure diffusion equations

To provide background, we first review multi-component diffusion in the absence of stress gradients. Under uniform temperature and pressure, gradients in endmember electrochemical potentials provide the driving force for diffusion of cations in ionic, crystalline solids (e.g., Yoo, 2020, eq. (7.9)). The electrochemical potential is the summation of a constituent's chemical potential and the electrostatic potential. The standard petrological diffusion model used to treat ionic, crystalline solids takes a mean field approach which assumes that the only driving force acting on a constituent is its electrochemical potential gradient (Lasaga, 1979). As such, the force can be written as:

$$F_I = \frac{\partial \mu_I}{\partial x} + z_I \mathcal{F} \frac{\partial \varphi}{\partial x} \qquad (1)$$

Where $F_I$ is the force acting on constituent $I$ (J mol$^{-1}$ m$^{-1}$), $\mu_I$ is the chemical potential (J mol$^{-1}$), $z_I$ is the charge number (unitless), $\mathcal{F}$ is the Faraday constant (Coulomb mol$^{-1}$), and $\varphi$ is the electrostatic potential (J Coulomb$^{-1}$). The constituent chemical potential gradient, $\frac{\partial \mu_I}{\partial x}$, can be replaced with the gradient in the associated endmember chemical potential. The chemical potential of an ion (e.g., Mg$^{2+}$) and its associated endmember (e.g., pyrope) are not numerically equivalent. However, their gradients are equivalent when: (1) the other constituents of the endmember are uniform and comprise a stable network structure that does not change appreciably during diffusion (i.e., the Al–Si–O framework in garnet; Lasaga, 1979) and (2) electroneutrality is maintained (Kroger, 1980; Schmalzried, 1995, p. 186-190, 205).

Assuming that the flux of some constituent $I$ is linearly proportional to its mobility and concentration together with the requirement of electrical neutrality (see Wendt, 1965), the 1-D relationship between the flux of cations and their associated endmember chemical potential gradients for ionic, multi-component crystalline solids can be defined as follows (equation (9) of Lasaga, 1979):

$$J_I = -u_I c_I \frac{\partial \mu_I}{\partial x} + u_I z_I c_I \frac{\sum_{K=1}^{N} z_K u_K c_K \frac{\partial \mu_K}{\partial x}}{\sum_{J=1}^{N} z_J^2 c_J u_J} \qquad (2)$$

where $J_I$ is the flux of the concentration of constituent $I$ (mol m$^{-3}$ m$^{-2}$ s$^{-1}$), $u_I$ is the mobility of constituent $I$ (m$^2$ mol J$^{-1}$ s$^{-1}$), $c_I$ is the concentration of constituent $I$ (mol m$^{-3}$), $z_I$ is the charge number (unitless), and $\mu_I$ is the endmember chemical potential per unit volume (J mol$^{-1}$ m$^{-3}$).

Equation (2) is appropriate for modeling diffusional fluxes in an ionic crystalline solid under uniform temperature and pressure conditions. The first term on the righthand side represents the flux due to the constituent's own chemical potential gradient, and the second term represents the flux due to the electrostatic potential gradient. The mobility of a constituent, $u_I$, defines its velocity relative to the applied thermodynamic driving force (e.g., equation (1)). It can be calculated from the experimentally measured self-diffusion coefficient of a constituent via the Nernst-Einstein equation

$$u_I = \frac{D_I}{RT} \qquad (3)$$

where $D_I$ is the self-diffusion coefficient (m$^2$ s$^{-1}$), $R$ is the ideal gas constant (J mol$^{-1}$ K$^{-1}$), and $T$ is the absolute temperature (K).

However, equation (2) requires two additional constraints for practical applications. Because a crystalline solid has a lattice with fixed sites, an $N$ component solid solution only has $N - 1$ independent chemical potential gradients. The $N^{th}$ dependent endmember is typically eliminated by employing the Gibbs-Duhem equation at a uniform temperature and pressure (e.g., Loomis, 1978, eq. (12); Lasaga, 1979, p. 456; Ganguly, 2002, p. 278; Yoo, 2020, eq. (4.35)) together with the additional constraint of electroneutrality:

$$\sum_{K=1}^{N} z_K c_K \frac{\partial \mu_K}{\partial x} = 0 \quad (4)$$

Equation (4) indicates that the net thermodynamic driving force is zero everywhere which also satisfies the requirements of crystallinity (Cahn and Larché, 1983; Gurtin et al., eq. (72.5)). Combining equations (2) and (4) yields:

$$J_I = -u_I c_I \frac{\partial \mu_I}{\partial x} + u_I z_I c_I \frac{\sum_{K=1}^{N-1} z_K c_K (u_K - u_N) \frac{\partial \mu_K}{\partial x}}{\sum_{J=1}^{N} z_J^2 c_J u_J} \quad (5)$$

Finally, the cation fluxes can be cast in terms of their associated endmember mole fractions (moles of $I$ per total moles) instead of concentrations (moles per m$^3$) by dividing equation (5) by the sum of all the endmember concentrations, $\sum_{K=1}^{N} c_K$. This is valid if the molar volume and, thus, the area over which the flux occurs does not change appreciably with diffusion, giving:

$$J_I^X = -u_I X_I \frac{\partial \mu_I}{\partial x} + u_I z_I X_I \frac{\sum_{K=1}^{N-1} z_K X_K (u_K - u_N) \frac{\partial \mu_K}{\partial x}}{\sum_{J=1}^{N} z_J^2 X_J u_J} \quad (6)$$

Where $J_I^X$ is the flux of the mole fraction of constituent $I$ (moles of $I$ per total moles m$^{-2}$ s$^{-1}$) and $X_I$ is the mole fraction of constituent $I$ (moles of $I$ per total moles).

Equation (6) defines the mole fraction flux of constituent $I$ counterbalanced by the arbitrarily chosen $N^{th}$ constituent via the Gibbs-Duhem equation to maintain charge neutrality and conserve lattice sites. We emphasize that it is strictly applicable only when there are no gradients in temperature and stress. The endmember chemical potentials can be calculated using any desired ideal or non-ideal activity model. Since chemical potential is a function of mole fraction, equation (6) is commonly re-written in terms of mole fraction gradients (e.g., Borinski et al., 2012, eq. (1–2)).

## 2.2. Diffusion with stress variation

When there are gradients in stress, the chemical potential gradient approach reviewed in Section 2.1 is insufficient for modeling intracrystalline diffusion because it is derived for uniform pressure conditions (Gibbs, 1878; Larché and Cahn, 1982, 1985). Instead, we use the relative chemical potential as it is defined under conditions of non-hydrostatic stress (Larché and Cahn, 1973, 1978b, 1982, 1985). Diffusion occurs while the exchange of constituents will continue to lower the energy locally. Equilibrium is achieved within a solid when no further exchanges can reduce its free energy, at which point all relative chemical potentials (rather than chemical potentials) are uniform.

The relative chemical potential between two constituents $I$ and $K$ at a given stress and temperature is given by the Larché-Cahn equation (Larché and Cahn, 1985, equation (4.21)). We write the equation following the terminology of Hess et al. (2022, equation (2)):

$$\mu_{I-K}(\sigma_{ij}, T) = \mu_I^*(P,T) - \mu_K^*(P,T) + RT \ln\left(\frac{a_I}{a_K}\right) - V_0 \eta_{ij}^{I-K}(\sigma_{ij} + P\delta_{ij}) - \frac{V_0}{2}\frac{\partial S_{ijkl}}{\partial X_{I-K}}(\sigma_{ij}\sigma_{kl} - P^2\delta_{ij}\delta_{kl}) \quad (7)$$

Where $\mu_{I-K}$ is the relative chemical potential (J mol$^{-1}$) at the given stress, $\sigma_{ij}$ (Pa), and temperature, $T$ (K$^{-1}$), $\mu_I^*$ is the chemical potential of the pure endmember $I$ (J mol$^{-1}$) at the reference pressure, $P$ (Pa), and temperature, $R$ is the ideal gas constant (J K$^{-1}$ mol$^{-1}$), $a$ is the activity, $V_0$ is the molar volume of the solid at the reference temperature and pressure (m$^3$ mol$^{-1}$), $\eta_{ij}^{I-K}$ is the partial molar strain tensor between $I$ and $K$ (unitless), $S_{ijkl}$ is the compliance tensor (Pa$^{-1}$), $X_{I-K}$ is the mole fraction of $I$ assuming dependent constituent $K$, and $\delta_{ij}$ is the Kronecker delta. See Hess et al. (2022) for details on solving the Larché-Cahn equation.

Equation (7) can be used to quantify the free energy change in a solid due to cation fluxes. In a system of $N$ endmembers, there are $N - 1$ unique, independent relative chemical potentials. If a solid is structurally isotropic, the formula for these relative chemical potentials simplifies considerably (Larché and Cahn, 1985; Hess et al., 2022):

$$\mu_{I-K}(\sigma_m, T) = \mu_I(\sigma_m, T) - \mu_K(\sigma_m, T) \tag{8}$$

where $\sigma_m$ is the mean stress (i.e., average of the three principal stresses) and $\mu$ is the endmember chemical potential in the solid solution. The endmember chemical potential is, for example, defined as: $\mu_I(\sigma_m, T) = \mu_I^*(\sigma_m, T) + RT \ln(a_I)$ for endmember $I$.

Crystalline solids, however, are not mechanically isotropic. Even minerals with cubic symmetry – the highest possible – are mechanically anisotropic (Nye, 1957; Cahn, 1962; Larché and Cahn, 1978a). This leads to anisotropic diffusion when a crystal is not at a uniform pressure (Larché and Cahn, 1982, 1985), including for garnet which is our focus herein. Nonetheless, the anisotropy of garnet elastic constants is very small (Erba et al., 2014). Consequently, any diffusional anisotropy in response to a stress gradient would be minimal. As such, we will treat garnet as isotropic as done in previous studies (e.g., Wheeler, 2018; Hess et al., 2022). Anisotropy is examined further in the discussion section.

Finally, we note that because vacancies mediate diffusion, they can be treated explicitly as a diffusing constituent (e.g., Smigelskas and Kirkendall, 1947; Darken, 1948; Mrowec, 1980; Chakraborty and Ganguly, 1991; Schmalzried, 1995; Mehrer, 2007; Li et al., 2018). However, vacancy mole fractions and chemical potentials are generally difficult to quantify accurately. Furthermore, treating vacancies explicitly in an ionic solid would require defining the stoichiometry and chemical potential of an associated neutral endmember which facilitates the diffusional exchange (e.g., Chakraborty and Ganguly, 1991). Given these complexities, it is most straightforward and standard practice to treat vacancies implicitly (Lasaga, 1979; Larché and Cahn, 1982, 1985).

## 2.3. Diffusional flux using relative chemical potential gradients

Larché and Cahn (1982, 1985) show that diffusion in crystalline network solids follows gradients in relative chemical potential (i.e., the spatial derivative of equation (7) or (8)). Using gradients in relative chemical potential, we can derive a new expression for diffusional flux.

First, we make a mean field assumption analogous to that in equation (1). That is, the forces acting on a constituent are its own *relative* endmember chemical potential gradient with an arbitrarily chosen $N^{th}$ endmember (e.g., Larché and Cahn, 1982; Cahn and Larché, 1983) and the electrostatic potential in the ionic solid. We further stipulate that the electrostatic potential is not a direct function of stress. Thus, we have:

$$J_I = -u_I c_I \left( \frac{\partial \mu_{I-N}}{\partial x} + z_I \mathcal{F} \frac{\partial \varphi}{\partial x} \right) \qquad (9)$$

Next, the requirements of electroneutrality indicate that the sum of the fluxes multiplied by the electrical charges of the diffusing constituents must be zero:

$$\sum_{K=1}^{N} z_K J_K = 0 \qquad (10)$$

When equation (9) is substituted into equation (10), the result is:

$$-\sum_{K=1}^{N} \left( u_K c_K z_K \frac{\partial \mu_{K-N}}{\partial x} + u_K c_K z_K^2 \mathcal{F} \frac{\partial \varphi}{\partial x} \right) = 0$$

which can be rearranged to solve for the electrostatic potential:

$$\mathcal{F} \frac{\partial \varphi}{\partial x} = -\frac{\sum_{K=1}^{N} \left( u_K c_K z_K \frac{\partial \mu_{K-N}}{\partial x} \right)}{\sum_{J=1}^{N} \left( u_J c_J z_J^2 \right)} \qquad (11)$$

When equation (11) is substituted back into equation (9), an expression for flux due solely to gradients in endmember relative chemical potentials is obtained:

$$J_I = -u_I c_I \frac{\partial \mu_{I-N}}{\partial x} + u_I c_I z_I \frac{\sum_{K=1}^{N}\left(u_K c_K z_K \frac{\partial \mu_{K-N}}{\partial x}\right)}{\sum_{J=1}^{N}(u_J c_J z_J^2)} \quad (12)$$

Equation (12) is directly analogous to equation (2), except that relative chemical potentials are used instead of chemical potentials. As a result, equation (12) treats diffusion under non-hydrostatic conditions, whereas equation (2) does not.

Two simplifications can be made to equation (12). First, concentration can be converted to mole fraction by dividing equation (12) by $\sum_{K=1}^{N} c_K$. Second, as the definition of relative chemical potential indicates that $\frac{\partial \mu_{N-N}}{\partial x} = 0$ (Larché and Cahn, 1973), the $N^{\text{th}}$ relative chemical potential gradient can be eliminated from equation (12) to give the result:

For $I = 1$ to $N - 1$:

$$J_I^X = -u_I X_I \frac{\partial \mu_{I-N}}{\partial x} + u_I X_I z_I \frac{\sum_{K=1}^{N-1}\left(u_K X_K z_K \frac{\partial \mu_{K-N}}{\partial x}\right)}{\sum_{J=1}^{N}(u_J X_J z_J^2)} \quad (13a)$$

And for $I = N$:

$$J_N^X = u_N X_N z_N \frac{\sum_{K=1}^{N-1}\left(u_K X_K z_K \frac{\partial \mu_{K-N}}{\partial x}\right)}{\sum_{J=1}^{N}(u_J X_J z_J^2)} \quad (13b)$$

Equation (13) defines the 1-D flux of the mole fraction of constituent $I$ ($J_I^X$; moles of $I$ per total moles m$^{-2}$ s$^{-1}$) as a function of relative chemical potential gradients under arbitrary stress conditions in an ionic, network solid. Since $N - 1$ endmembers are independent, only the first $N - 1$ equations given by equation (13a) are necessary because the flux of the $N^{\text{th}}$ endmember is the difference of one minus the sum of the other $N - 1$ endmember mole fractions. Nonetheless, equation (13b) explicitly allows for the calculation of the flux of the $N^{\text{th}}$ endmember, providing a means to test that the lattice constraint is indeed being satisfied by confirming that the sum of the fluxes is zero.

The form of equation (13) is similar to the flux equation based on chemical potential gradients (equation (6)). The first term of equation (13a) represents the flux due to the constituent's own relative chemical potential gradient. The second term represents the flux due to the electrostatic potential gradient in the ionic solid. Unlike equation (6), however, equation (13) is valid under conditions of non-hydrostatic stress as it is written as a function of relative chemical potential gradients. Equation (4) is only applicable when pressure and temperature are uniform, and therefore, equation (6) is invalid when stress gradients exist. It is worth emphasizing that equation (13) is appropriate to model diffusion in network solids with heterogeneous stresses because the Gibbs-Duhem equation is never applied in its derivation.

The terms in equation (13) can be determined as follows: The mobilities, $u_I$, are calculated from self-diffusion coefficients using equation (3) (e.g., Larché and Cahn, 1982; Larché and Voorhees, 1996). The charge number, $z_I$, is the unitless charge value of the diffusing cation associated with each endmember (e.g., +2 for $Mg^{2+}$ in pyrope garnet). Relative chemical potential can be calculated using the Larché-Cahn equation (equation (7)). This equation indicates that crystallographic and stress orientations need to be considered when computing relative chemical potential values. However, the much simpler equation (8) is applicable if a mineral can reasonably be approximated as isotropic (e.g., garnet; see section 4.2.). Equation (13) is valid for any desired activity model (ideal or non-ideal), depending on how one determines the relative chemical potentials. The choice of the $N^{th}$ endmember is arbitrary (Larché and Cahn, 1985). If equation (13) is extended beyond one dimension, care must be taken to account for the effects of orientation on both self-diffusion coefficients and relative chemical potentials (Larché and Cahn, 1973, 1985; Wheeler, 2018; Hess et al., 2022).

Diffusion following relative chemical potential gradients may be written as a linear combination of mobility-based interdiffusion coefficients and relative chemical potential gradients (e.g., Larché and Cahn, 1985, equation (8.1)):

$$J_I^X = -\sum_{J=1}^{N-1} B_{IJ} \frac{\partial \mu_{J-N}}{\partial x} \tag{14}$$

When equation (13a) is rewritten in the form of equation (14), then:

$$B_{IJ} = u_I X_I \delta_{IJ} - u_I X_I z_I \frac{u_J X_J z_J}{\sum_{K=1}^{N} u_K X_K z_K^2} \tag{15}$$

Where $\delta_{IJ}$ is the Kronecker delta, $B_{IJ}$ is the relative chemical potential-based interdiffusion coefficient between constituents $I$ and $J$ (m² mol J⁻¹ s⁻¹), and all other terms are as previously defined. We use $B_{IJ}$ following Larché and Cahn (1982) in place of the more familiar interdiffusion coefficient notation of $D_{IJ}$ to emphasize that $B_{IJ}$ represents interdiffusion coefficients that are calculated for use with relative chemical potential gradients.

Importantly, the mobility-based interdiffusion coefficient matrix, $B_{IJ}$, is symmetric, upholding the fundamental Onsager reciprocal relations (ORRs; Onsager, 1931). This is easily shown by comparing $B_{IJ}$ and $B_{JI}$ for the case of $I \neq J$:

$$-u_I X_I z_I \frac{u_J X_J z_J}{\sum_{K=1}^{N} u_K X_K z_K^2} = -u_J X_J z_J \frac{u_I X_I z_I}{\sum_{K=1}^{N} u_K X_K z_K^2}$$

Clearly, $B_{IJ}$ (left side) is equal to $B_{JI}$ (right side). Given the additional complexities of crystalline solids compared to fluids (e.g., crystal symmetries and the ability to support non-hydrostatic elastic stresses at mechanical equilibrium), verification of the ORRs remains difficult and is not guaranteed in all reference frames for solids (e.g., molar vs. barycentric reference frame; Mullins and Sekerka, 1981; Cahn and Larché, 1983). Thus, the symmetry of $B_{IJ}$ (eq. (15)) offers an important check for the validity of our derivation based on relative chemical potential gradients. We also note that while the standard flux expression (eq. (6)) yields asymmetric interdiffusion coefficient matrices, this does not invalidate the expression for use at uniform pressure and temperature conditions. Equation (6) uses chemical potential gradients which are not independent driving forces as shown in equation (4). The interdependency of the gradients yields asymmetric interdiffusion coefficient matrices without necessarily violating the ORRs (Cahn and Larché, 1983).

Application of equation (15) to, for example, a four-component system such as almandine–pyrope–grossular–spessartine (Alm–Py–Grs–Sps) garnet generates a 3 × 3 matrix of relative chemical potential-based interdiffusion coefficients that can be used with the three arbitrarily chosen independent relative chemical potentials. Because all substitutional cations have the same +2 charge, the $z$ terms cancel.

In summary, equations (13) and (15) provide a new 1-D formulation that can be used for calculating diffusion in ionic, crystalline solids with any symmetry and under arbitrary stress conditions. Herein, we apply these expressions to Alm–Py–Grs–Sps garnet. We use the Holland and Powell (2011) thermodynamic dataset and their equation of state to calculate pure endmember chemical potentials. We calculate garnet endmember activities following White et al. (2014) and self-diffusion coefficients using Chakraborty and Ganguly (1992) and Faryad and Chakraborty (2005). In models where pressure is not uniform, the self-diffusion coefficients vary as a function of mean stress (pressure). We obtain similar results with other garnet activity models (e.g., Berman, 1990; Ganguly et al., 1996) and diffusion coefficients (e.g., Carlson, 2006; Chu and Ague, 2015).

We model diffusion using a 1-D spherical diffusive transport code that employs the forward time centered space (FTCS) method (e.g., Roache, 1972). We use a zero-flux boundary condition at the garnet rims. This means there is no exchange with the matrix which allows us to isolate the effects of internal stress variation in the crystal. The interdiffusion coefficients are re-calculated at every time step as composition changes. We use the initial condition (composition, stress, and strain) as our reference frame. Thus, any changes in composition, stress, or strain in the solid are mapped onto the location of that unit volume in the reference frame (Larché and Cahn, 1985).

We assume that garnet behaves perfectly elastically and that the stress gradient does not change with time in the examples below. This simplification allows us to highlight the effects of stress on multi-component mineral composition and to compare our results to the equilibrium conditions of Larché and Cahn (1973, 1985). In addition, while we explore how an imposed stress gradient affects diffusion, cation diffusion may itself generate stresses ("self-stress") due to differences in atomic sizes (e.g., Larché and Cahn, 1982; Schmalzried, 1995, p. 71; Van Orman et al., 2001; Cherniak and Watson, 2003; Baumgartner et al., 2010; Zhong et al., 2017). Nonetheless, we emphasize that equation (13) is general and is not limited to the elastic stresses or constant imposed stress gradients considered herein. Although both viscous relaxation (i.e., solid-state creep) and diffusion-induced self-stresses are beyond the scope of our present work, they represent important processes that can be investigated using our derivation (equation 13) in future studies.

## 3. Results

Equation (13) allows diffusion to be treated in ionic, multi-component minerals when pressure is not uniform. Here, we first show that diffusion modeled using relative chemical potential gradients (equation (13)) at uniform pressure and temperature conditions provides diffusion profiles (and thus, diffusion rates) identical to standard methods based on chemical potential or mole fraction gradients (equation (6); see also Borinski et al., 2012, equations (1–2)). Second, we demonstrate that relative chemical potential gradients are valid for modeling diffusion when there are stress gradients, whereas mole fraction and chemical potential gradients are not. Third, we present expected behaviors for stress-induced diffusion in garnet and show that they match equilibrium compositions predicted by Larché-Cahn theory (Larché and Cahn, 1985).

Under uniform temperature and pressure conditions, the derivations of Larché and Cahn (1973, 1985) reduce to the equilibrium conditions of Gibbs (1878). Therefore, diffusion modeled using relative chemical potential gradients with a uniform pressure provides equivalent results to diffusion modeled using chemical potential gradients or mole fraction gradients. Figure 1 shows a comparison between the modeled profile using relative chemical potential gradients (equation (13)) and chemical potential or mole fraction gradients (equation (6)) for a pyrope-rich garnet composition. The initial composition is a step function (thin dashed lines) which relaxes for two million years at 700 °C and 1 GPa. The results for the two types of potential gradients are identical, as expected. Importantly, this also indicates that the diffusion rates calculated using equations (13) and (15) are the same as those computed using standard methods. Thus, the results obtained from our derivation can be directly compared to other works when pressure is uniform.

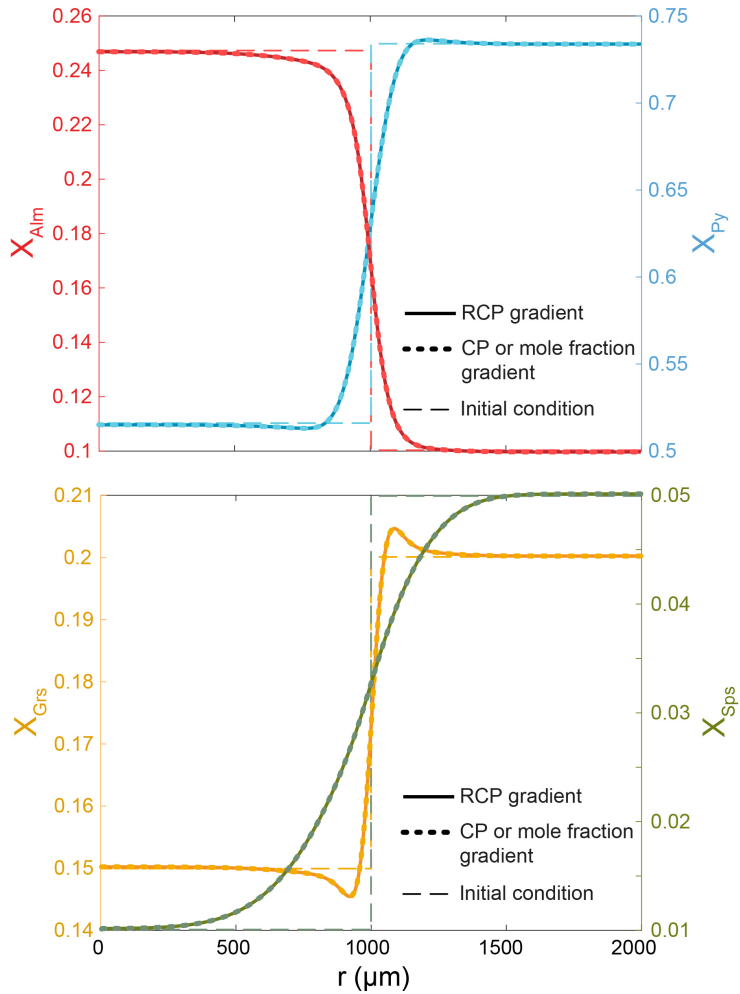

Figure 1: Diffusion in garnet following relative chemical potential gradients (RCP, solid lines) and chemical potential (CP) or mole fraction gradients (dashed lines). A spherical garnet composed of almandine (Alm), pyrope (Py), grossular (Grs), and spessartine (Sps) has a radius of 2,000 μm and an internal step function in composition (thin dashed lines). Diffusion is simulated for 2 million years at 700 °C and 1 GPa. The activities are calculated following White et al. (2014). The self-diffusion coefficients are $D_{Alm}$ = 5.29 x $10^{-23}$ m$^2$ s$^{-1}$, $D_{Py}$ = 3.07 x $10^{-23}$ m$^2$ s$^{-1}$, $D_{Grs}$ = 2.64 x $10^{-23}$ m$^2$ s$^{-1}$, and $D_{Sps}$ = 6.07 x $10^{-22}$ m$^2$ s$^{-1}$.

In contrast, when there is a stress gradient, the equivalency between diffusion models that use mole fraction, chemical potential, or relative chemical potential gradients ceases (Fig. 2). For example, suppose that a spherical garnet with an initially homogeneous composition is subjected to a constant, hypothetical internal stress gradient as shown in Figure 2a. We use a large variation in mean stress to highlight the interactions between stress and diffusion more clearly.

The garnet will have higher strain energy at its core than its rim, and consequently, the strain energy gradient will drive diffusion.

However, if mole fraction gradients are used to model diffusion, no diffusion is predicted because there is no initial gradient in mole fraction (Fig. 2b). When equation (6) is re-written in terms of mole fraction (e.g., Borinski et al., 2012), it is assumed that the chemical potential is only a function of mole fraction, and therefore any pressure or temperature variations are ignored.

On the other hand, when chemical potential gradients are used, the effects of stress on the endmember chemical potentials are included, but in an incorrect way. As garnet is a crystalline solid, the net flux must be zero everywhere to maintain the lattice (e.g., Yoo, 2020, p. 163). At uniform pressure, the Gibbs-Duhem equation (equation (4)) accomplishes this by ensuring the net driving force (and therefore flux) is zero. However, when pressure is not uniform, equation (4) is unjustified which invalidates equation (6). Consequently, the $N^{th}$ constituent (in this case spessartine, arbitrarily chosen) is forced to balance the flux in an unrealistic manner. In the presence of a gradient in internal mean stress as in Figure 2a, every constituent will have a higher endmember chemical potential at the core of the garnet than the rim. As such, without the lattice constraint, every cation would diffuse toward the rim. However, since the diffusing cation of the $N^{th}$ endmember ($Mn^{2+}$) provides a counter-flux to maintain the crystal lattice, the mole fraction of spessartine quickly becomes negative in the rim (Fig. 2c). As such, the simulation ends at 0.2 My after which point the model becomes unstable and fails. Thus, neither diffusion using mole fraction nor chemical potential gradients provides realistic results when there is a stress gradient.

Instead, relative chemical potential gradients must be used (Larché and Cahn, 1982, 1985; Cahn and Larché, 1983; Balluffi et al., 2005, p. 42–43; Gurtin et al., 2010, p. 398–400).

The relative chemical potential quantifies energy changes due to the exchange of constituents on lattice sites. Since the lattice constraint is satisfied without the Gibbs-Duhem equation (Cahn and Larché, 1983), plausible diffusion is predicted (Fig. 2d). Consequently, equation (13) can be used to model flux when stress gradients exist. The mole fraction of pyrope is predicted to increase toward the high stress region while the mole fractions of grossular and spessartine increase toward the low stress region. The mole fraction of almandine changes by only a very minor amount for this composition (Fig. 2d).

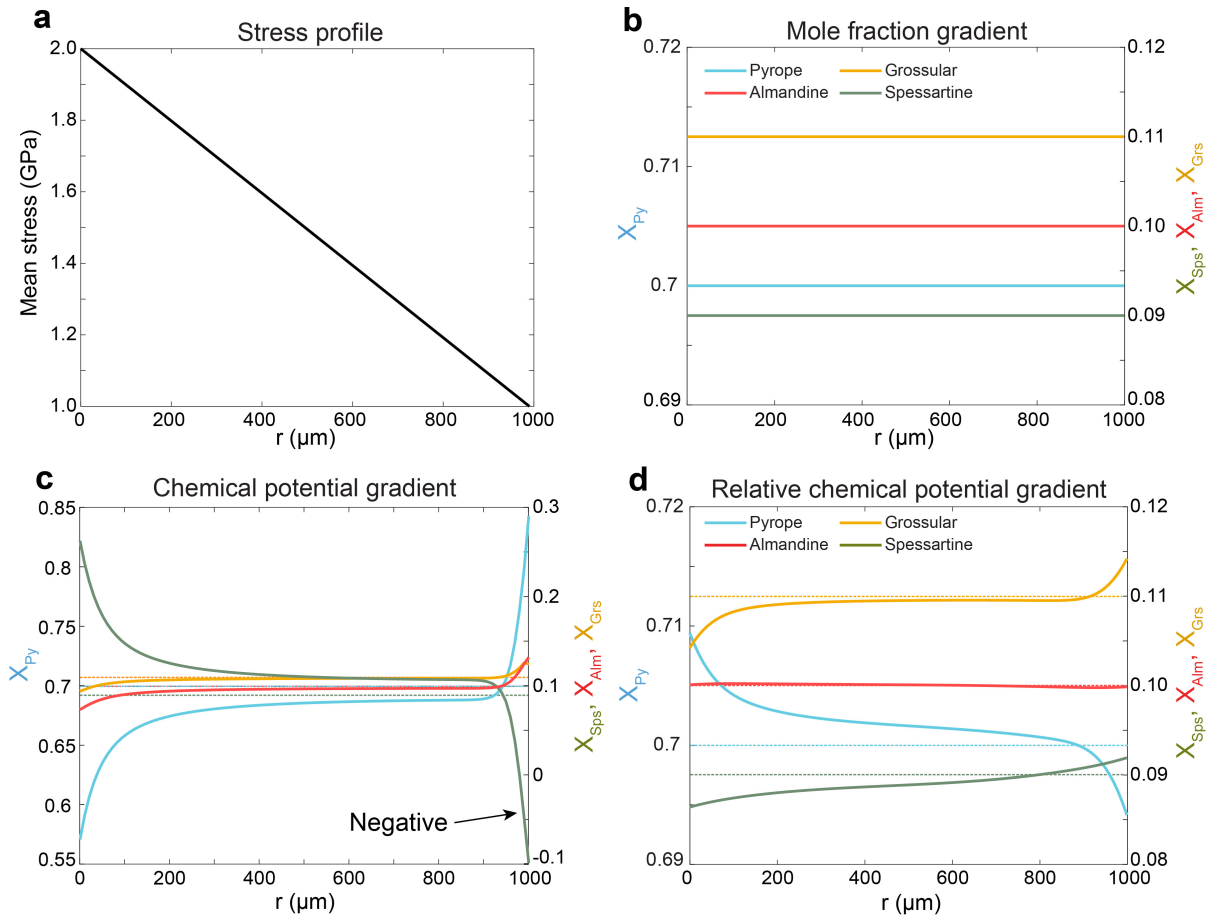

Figure 2: Comparing simulated diffusion profiles at 700 °C in a stressed, pyrope-rich garnet. a) The stress profile of the garnet. Diffusion profiles are simulated using the following driving forces: b) chemical potential-based mole fraction gradients for 5 My, c) chemical potential gradients for 0.2 My, and d) relative chemical potential gradients for 5 My. Thin dotted lines represent the initial composition. The self-diffusion coefficients are a function of mean stress. The ranges are $D_{Alm}$ = 2.65 x $10^{-23}$ to 5.29 x $10^{-23}$ m² s⁻¹, $D_{Py}$ = 1.59 x $10^{-23}$ to 3.07 x $10^{-23}$ m² s⁻¹, $D_{Grs}$ = 1.32 x $10^{-23}$ to 2.64 x $10^{-23}$ m² s⁻¹, and $D_{Sps}$ = 2.89 x $10^{-22}$ to 6.07 x $10^{-22}$ m² s⁻¹.

Figure 3 represents a relative chemical potential-based diffusion profile with the same internal stress profile and initial condition as in Figure 2 but at a temperature of 800 °C and a time of ten million years such that the internal garnet composition is near equilibrium. In addition, both ideal (dashed lines) and non-ideal (solid lines) mixing behaviors are compared. One GPa of mean stress variation leads to a 0.06–0.08 variation in $X_{Py}$ and $X_{Grs}$ from core to rim with a much smaller variation in $X_{Alm}$ and $X_{Sps}$. This suggests that a large mean stress variation (>100 MPa) is required for there to be a discernable change in mineral composition.

Figure 3 also highlights the importance of incorporating non-ideal mixing behavior (Hess et al., 2022). Non-ideality increases the modeled composition change of all endmembers by up to a factor of two under the conditions of Figure 3. Thus, when modeling stress-induced diffusion, non-ideal mixing behavior should be considered.

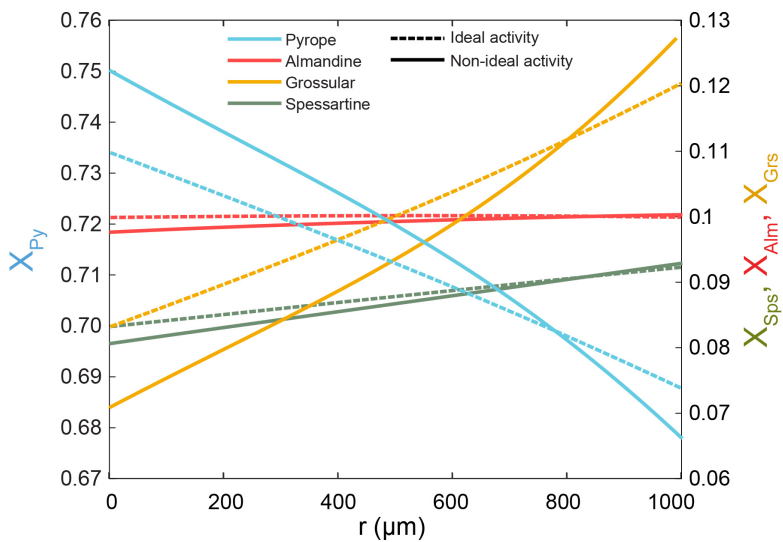

Figure 3: Comparing diffusion profiles predicted for an internal stress gradient in garnet with non-ideal (White et al., 2014) and ideal activity models using relative chemical potential gradients. The temperature is 800 °C and the time is 10 My. The stress profile and other model parameters are the same as in Figure 2a. The self-diffusion coefficients are a function of mean stress. The ranges are $D_{Alm}$ = 7.18 x $10^{-22}$ to 1.34 x $10^{-21}$ m$^2$ s$^{-1}$, $D_{Py}$ = 4.76 x $10^{-22}$ to 8.63 x $10^{-22}$ m$^2$ s$^{-1}$, $D_{Grs}$ = 3.59 x $10^{-22}$ to 6.72 x $10^{-22}$ m$^2$ s$^{-1}$, and $D_{Sps}$ = 6.15 x $10^{-21}$ to 1.20 x $10^{-20}$ m$^2$ s$^{-1}$.

Figure 4 shows the temporal evolution of the composition of a more almandine-rich garnet (Fig. 4a) as well as the evolution of the relative chemical potentials (Fig. 4b). The stress conditions are identical to the profile shown in Figure 2a, and the composition is initially homogeneous. As time progresses, the mole fractions of each endmember change from initially uniform, flat lines to curved lines that vary by up to several hundredths of a mole fraction from core to rim (Fig. 4a). On the other hand, the relative chemical potential profiles have the opposite behavior (Fig. 4b). They begin as initially curved lines, but after equilibrium is achieved, they are uniform and flat. This demonstrates the important result that the underlying relative chemical potential is the indicator of whether equilibrium has been attained. Thus, unlike in standard diffusion models, equilibrium is not necessarily achieved when the composition is homogeneous (Larché and Cahn, 1982, 1985; Zhong et al., 2017).

When the relative chemical potentials are uniform throughout the garnet, the free energy of the garnet cannot be lowered by further diffusion and equilibrium is achieved (Larché and Cahn, 1982, 1985). At this point, the composition is identical to the equilibrium composition predicted by Larché-Cahn theory (Larché and Cahn, 1973, 1985; Powell et al., 2018; Wheeler, 2018; Hess et al., 2022). This serves as a test of both our numerical algorithm and solution procedure for the transient problem.

In summary, we show that our diffusional flux expression (equation (13)): (1) provides diffusion rates that are identical to those obtained using standard methods when pressure is uniform (Fig. 1), (2) can incorporate stress gradients (Fig. 2), and (3) leads to composition profiles with uniform underlying relative chemical potentials when stress gradients are present as predicted by Larché-Cahn theory (Fig. 4).

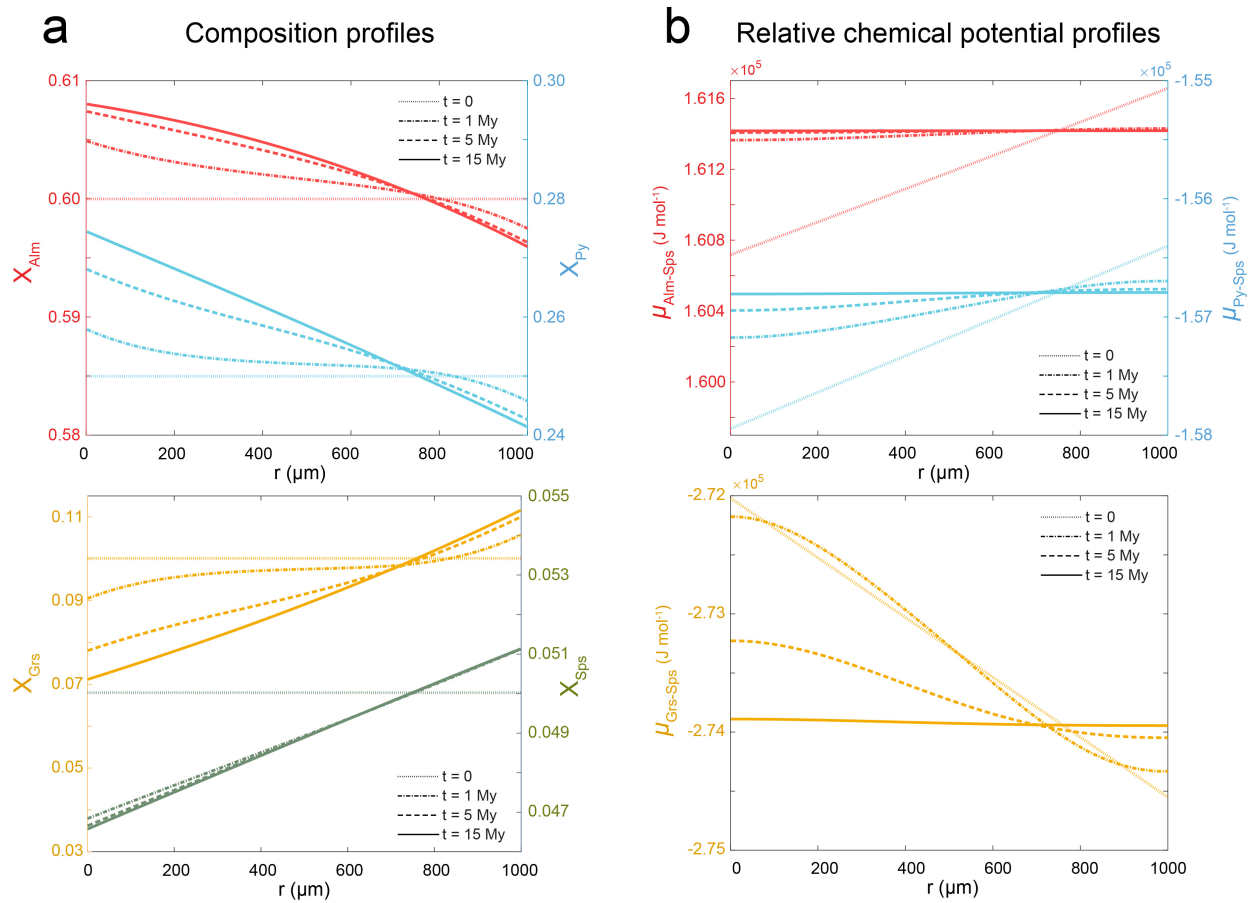

Figure 4: Temporal evolution of an almandine-rich garnet with an internal stress gradient as shown in Figure 2a. The self-diffusion coefficients are identical to those in Figure 3 and the temperature is 800 °C. a) Composition profiles as a function of time. b) Relative chemical potential profiles a function of time. Note that four endmembers give rise to three unique relative chemical potentials.

## 4. Discussion

### 4.1. Interactions between stress and composition

When there are gradients in stress, new diffusion behaviors arise that are not predicted by the standard, uniform pressure models such as equation (6). When temperature and pressure are uniform, chemical potentials and relative chemical potentials are only a function of composition.

However, stress variation creates a gradient in strain energy which also affects relative chemical potential. Strain energy is partially a function of endmember molar volumes or lattice parameters. Consequently, cations associated with the largest endmembers will diffuse toward low-stress regions and the cations associated with the smallest endmembers will diffuse toward high-stress regions to diminish the strain energy gradient (Tajčmanová et al., 2015, 2021; Hess et al., 2022). This phenomenon is known as the Gorsky effect (Gorsky, 1935) and is commonly observed, for example, in interstitial hydrogen diffusion in stressed alloys (e.g., Fukai et al., 1985; Pálsson et al., 2012; Shi et al., 2018).

Equilibrium is achieved when the change in compositional energy and the change in strain energy from any movement of constituents are equal. At this point, all endmember relative chemical potentials are uniform, and there is no thermodynamic driving force for further diffusion. The endmember mole fractions, and their associated relative chemical potentials, are then equal to the equilibrium compositions calculated via equation (7) (Larché and Cahn, 1985; Powell et al., 2018, 2019; Wheeler, 2018; Hess et al., 2022). As emphasized above, although the relative chemical potentials are uniform, the equilibrium mineral composition need not be (Fig. 4).

The composition changes predicted in Figures 3 and 4 are consistent with the Gorsky effect. The mole fraction of $Mg^{2+}$, associated with the smallest endmember (pyrope), increases toward the core of the garnet where the stress is highest. In contrast, the mole fraction of $Ca^{2+}$, associated with the largest endmember (grossular), increases toward the rim. Since these two cations and their associated endmembers have the largest molar volume difference, they represent most of the observed composition change. In contrast, $Fe^{2+}$ and $Mn^{2+}$, associated with almandine and spessartine, respectively, have endmember molar volumes between pyrope and grossular. Thus, $Fe^{2+}$ and $Mn^{2+}$ mole fractions do not change as much as $Mg^{2+}$ and $Ca^{2+}$.

Despite these general trends, the details of how cation flux and stress interact can vary considerably depending on the composition. For example, for the pyrope-rich composition, the mole fraction of $Fe^{2+}$ slightly decreases with increasing stress (Fig. 3). In contrast, the mole fraction of $Fe^{2+}$ increases with increasing stress in the example garnet with a more almandine-rich composition (Fig. 4). Thus, while the cations associated with the largest and smallest endmembers (grossular and pyrope, respectively) will have behaviors predicted simply by the Gorsky effect, the behaviors of the cations associated with intermediate endmembers can vary depending on composition.

Additionally, non-ideal mixing is important to consider (Hess et al., 2022). The interaction parameters of garnet endmembers have positive deviations from ideality (e.g., Berman, 1990; Ganguly et al., 1996; White et al., 2014). Consequently, mixing of cations becomes less energetically favorable relative to ideal mixing behaviors. Therefore, it takes a greater amount of composition change to balance the strain energy gradient imposed by a stress gradient (Fig. 3). Pyrope and grossular have a highly non-ideal mixing behavior, meaning that stress has a larger effect on the flux of $Mg^{2+}$ and $Ca^{2+}$. Almandine and spessartine endmembers, however, have smaller deviations from ideality with respect to all endmembers (e.g., White et al., 2014), and hence $Fe^{2+}$ and $Mn^{2+}$ have a more modest flux in response to gradients in stress.

Finally, we note that while stress drives diffusion, the diffusion of cations with different endmember molar volumes may also generate stresses which locally affect composition (Larché and Cahn, 1982, 1985; Van Orman et al., 2001; Cherniak and Watson, 2003; Baumgartner et al., 2010; Zhong et al., 2017). Diffusion-induced "self-stress" in a sphere can be quantified, for example, by using Larché and Cahn (1982)'s equation (26), and then its effects on relative chemical potential may be subsequently incorporated into our flux expression through equation (7). However, we deem this beyond our scope here. We make a limiting assumption that the stress profile remains constant (Fig. 2a) to highlight the expected direction and magnitudes of stress-induced diffusion in garnet (Fig. 2d, 3, 4). It is worth emphasizing, however, that this assumption is not required when applying equation (13) as it is perfectly general and can incorporate stresses from any source using equation (7).

## 4.2. Crystallographic anisotropy

We have applied our derived flux expression (equation (13)) to garnet which has cubic symmetry. Although cubic minerals are often thought of as isotropic, their mechanical properties (i.e., Young's moduli) are described by fourth-order tensors and are thus anisotropic (Nye, 1957). Consequently, the effect of stress on relative chemical potential will vary directionally, leading to anisotropic diffusion rates even in cubic minerals (Cahn, 1962; Larché and Cahn, 1982). Nonetheless, we treat garnet as isotropic because its mechanical anisotropy is very small (Erba et al., 2014). The spatial variation in diffusion rates due to this anisotropy would be only a few percent which is negligible compared to the uncertainties of analytical techniques and the diffusion coefficients themselves.

The effects of anisotropy in lower symmetry structures, however, will have a more appreciable effect on stress-induced diffusion. Non-cubic minerals have anisotropy in their lattice dimensions as well as mechanical properties. Both Wheeler (2018) and Hess et al. (2022) show that the mole fraction of albite in the mineral plagioclase, for example, can increase in response to stress in one crystallographic orientation and decrease in response to stress in another. Consequently, stress-induced diffusion in plagioclase would not only have anisotropic diffusion rates but would also have the direction of cation diffusion change as a function of crystallographic orientation. Cations associated with the albite endmember, for example, would diffuse toward high stress regions along one crystallographic axis and toward low stress regions along a different axis (Hess et al., 2022).

Thus, caution must be exercised when predicting the behaviors of anisotropic minerals without a full consideration of their lattice parameters and elastic moduli. For non-cubic minerals or cubic minerals with considerable mechanical anisotropy, the isotropic relative chemical potential approximation (i.e., equation (8)) should not be used. The full effects of crystallographic and stress tensor orientation on relative chemical potential should instead be incorporated using equation (7), even in the case of a simple 1-D planar diffusion model.

### 4.3. Potential Applications

A promising environment to test and apply simple cases of stress-driven diffusion models may be around garnet-hosted coesite inclusions (e.g., Chopin, 1984; Massonne, 2001; Lü and Zhang, 2012). The presence of preserved coesite or former coesite suggests GPa-level stress variations existed during exhumation to prevent the transition from coesite to quartz (e.g., Parkinson and Katayama, 1999; Parkinson, 2000). In addition, other mineral inclusions such as quartz and zircon in garnet are predicted to develop stresses due to differences in lattice parameters and elastic constants during exhumation (e.g., Guiraud and Powell, 2006; Kohn, 2014; Murri et al., 2018; Thomas and Spear, 2018; Mazzucchelli et al., 2019; Moulas et al., 2020; Gilio et al., 2021). Depending on the entrapment conditions, stress variations in the host of 0.5-1 GPa or greater are indicated (e.g., Thomas and Spear, 2018; Bonazzi et al., 2019; Zhong et al., 2019; Campomenosi et al., 2020). Preserved diffusion profiles around inclusions could then be used to estimate internal stresses and time scales of metamorphic events even if there is no initial chemical heterogeneity. However, as we have previously noted, large differences in mean stress (>100 MPa) are required to shift mole fractions by a few hundredths. Such small composition changes would only be revealed by very careful chemical microanalysis procedures.

It is also important to point out that stress variation doesn't necessarily equate to a mean stress gradient. For example, it is well known that an overpressured spherical inclusion inside of an elastic isotropic host will not create a mean stress gradient in the host (e.g., Timoshenko and Goodier, 1970). Nonetheless, non-spherical inclusion geometries (e.g., King et al., 1991; Moulas et al., 2014; Campomenosi et al., 2018, 2020; Moore et al., 2019; Zhong et al., 2021), interactions between multiple inclusions (Voorhees and Johnson, 2004), mechanical or crystallographic anisotropy in the inclusion(s) or host (King et al., 1991), and alternative stress models such as the multi-anvil model (Tajčmanová et al., 2014) all have the potential to create mean stress gradients that would drive diffusion in perfectly elastic crystals.

Furthermore, the assumption of elastic behavior is not necessary. Our examples in Figures 2–4 assume a constant gradient in elastic stress to allow for direct comparison with the equilibrium conditions of Larché and Cahn (1973, 1985) which require elasticity. However, since diffusion is inherently a non-equilibrium process, transient viscous behavior (i.e., solid-state creep) can be incorporated because the network model allows for dislocations to modify the structure (Larché and Cahn, 1985). As such, standard mechanical models are fully compatible with Larché-Cahn theory. While viscous relaxation would ultimately reduce stress gradients in garnet (e.g., Moulas et al., 2020; Zhong et al., 2020), viscous behavior can lead to transient GPa-level stresses which may drive significant diffusion (e.g., Zhang 1998; Moulas et al., 2014; Tajčmanová et al. 2014; Dabrowski et al. 2015; Zhong et al., 2017). Stephenson (1986, 1988) and Erdelyi and Schmitz (2012), for example, demonstrated how viscous behavior can be incorporated into diffusion models based on Larché-Cahn theory. Equation (13) can similarly be used as a foundation to investigate the effects of viscous behavior on diffusion in ionic, crystalline solids.

**4.4. Geochemical Implications**

Chemical zonation in minerals records unique information about evolving pressures, temperatures, and system chemistry in Earth's crust and mantle, thus providing fundamental insights into geochemical cycling, orogenesis, and myriad other Earth system phenomena. Diffusional relaxation of chemical heterogeneities gives crucial perspectives on the timescales of metamorphism, igneous activity, and hydrothermal circulation, as it modifies pre-existing zonation acquired via mineral growth, dissolution–precipitation, or other processes (e.g., Erambert and Austrheim, 1993; Penniston-Dorland, 2001; Cherniak and Watson, 2003; Watson and Baxter, 2007; Faryad and Chakraborty, 2005; Caddick et al., 2010; Ague and Carlson, 2013; Chu et al., 2017, 2018; Kohn and Penniston-Dorland, 2017; Tan et al., 2020; Zou et al., 2021). It has long been recognized that the chemical potential gradients that arise due to chemical zonation will drive diffusion. But internal stress variations within mineral grains associated with, for example, over- or underpressured mineral inclusions or larger scale deformational forces may also act to drive diffusion (e.g., Parkinson, 2000; Jamtveit et al., 2019; Moore et al., 2019; Campomenosi et al., 2020; Kaatz et al., 2021). Consequently, rigorous implementation of geothermometry, geobarometry, diffusion chronometry, and element partitioning studies should consider the potential impacts of stress on diffusional relaxation. We posit that these impacts can now be treated using the diffusion flux equation (13) derived herein, making it possible to quantitatively assess how stress influences the evolving chemical zonation of minerals.

## 5. Summary

We have derived an expression for modeling 1-D diffusional fluxes in ionic, crystalline solids under arbitrary stress states that is consistent with Larché-Cahn theory (Larché and Cahn, 1973, 1982, 1985). Using this expression, we modeled examples of stress-induced diffusion in the common quaternary garnet solid solution assuming elastic behavior and a constant internal stress gradient. Cation fluxes due to stress gradients are determined by the relative differences in endmember molar volumes or lattice parameters, by elastic moduli, and by the non-ideal activity interaction parameters.

We treat garnet as mechanically isotropic as it is cubic and only weakly mechanically anisotropic (Erba et al., 2014). However, in general, anisotropy in both lattice parameters and elastic moduli may have appreciable effects on both rates and directions of stress-induced diffusion. Anisotropy is especially important to incorporate when modeling stress-induced diffusion in minerals with lower symmetries even in simple 1-D planar diffusion models (Larché and Cahn, 1985; Wheeler, 2018; Hess et al., 2022).

In general, we have shown that stress-induced diffusion will not lead to a uniform composition, consistent with earlier studies (e.g., Larché and Cahn, 1982, 1985; Zhong et al., 2017). Instead, internal stress variations of a few hundred MPa or greater will result in endmember mole fraction variations on the order of several hundredths. Such large stress variations may develop, for example, around stressed inclusions with interacting stress fields (e.g., Voorhees and Johnson, 2004) or around individual geometrically anisotropic stressed inclusions (King et al., 1991). Stress-induced diffusion profiles may potentially provide time estimates for the duration of these intracrystalline stresses and related metamorphic events such as exhumation. Although stress-induced compositional effects will likely be subtle, they nonetheless could reveal substantial deviations from hydrostatic conditions.

Finally, we note that our derivation is sufficiently general as to be applied to more complex phenomena such as diffusion-induced stresses (e.g., Larché and Cahn, 1982) and viscous relaxation (e.g., Stephenson, 1986, 1988; Erdelyi and Schmitz, 2012). The ability to model diffusion under arbitrary stress conditions in a way that is consistent with Larché-Cahn theory presents new possibilities for extracting both stress and time information from mineral compositions which, in turn, can provide a deeper understanding of large-scale tectonic processes.

**Acknowledgements:**

We thank X. Chu, A. A. Haws, M. T. Brandon, and P. W. Voorhees for thoughtful discussions, L. Tajčmanová, X. Zhong, and an anonymous reviewer for their critical and insightful comments, and J. M. Brenan for editorial handling. Financial support from the US National Science Foundation Directorate of Geosciences (EAR-2208229) and Yale University are gratefully acknowledged.